\documentclass[5p]{elsarticle}

\usepackage{natbib}
\usepackage{amsmath}
\usepackage[binary-units,per-mode=symbol]{siunitx}
\usepackage{tikz}
\usepackage{listings}
\usepackage{pgfplotstable}
\usepackage{array}
\usepackage{booktabs}
\usetikzlibrary{shapes.misc}
\pgfplotstableset{
  every head row/.style={before row=\toprule, after row=\midrule},
  every last row/.style={after row=\bottomrule}
}

\journal{Astronomy and Computing}
\biboptions{authoryear}

\begin{document}

\begin{frontmatter}

\title{Faster GPU-based convolutional gridding via thread coarsening}

\author[mymainaddress]{Bruce Merry}
\ead{bmerry@ska.ac.za}
\address[mymainaddress]{SKA South Africa, 3rd Floor, The Park, Park Road, 7405 South Africa}

\begin{abstract}
Convolutional gridding is a processor-intensive step in interferometric
imaging. While it is possible to use graphics processing units (GPUs) to
accelerate this operation, existing methods use only a fraction of the
available flops. We apply thread coarsening to improve the
efficiency of an existing algorithm, and observe performance gains of up to
$3.2\times$ for single-polarization gridding and $1.9\times$ for
quad-polarization gridding on a GeForce GTX 980, and smaller but still
significant gains on a Radeon R9 290X.
\end{abstract}


\end{frontmatter}


\section{Introduction}
Interferometric imaging is a key tool in radio astronomy, but as modern
instruments provide more antennas, longer baselines, and more channels, it is
becoming increasingly computationally costly. A major component of an imaging
pipeline is \emph{convolutional gridding}, as well as the corresponding
degridding for predicting visibilities.

Given the computational cost of gridding, it is natural to apply
accelerator hardware, of which the cheapest and most ubiquitous is the
Graphics Processing Unit (GPU). However, the irregular data access patterns
make this a non-trivial task. One of the first really practical algorithms for
GPU-accelerated gridding is due to \citet{romein-gridding}. Despite being state of
the art, it typically spends only about 25\% of a GPU's compute power on the
actual convolution operations. There are bottlenecks in the memory system, but
also computational overheads associated with address calculations. Our goal is
to reduce these overheads to make more flops available for the convolution
calculations.

Our contribution is a modification to the algorithm in which each thread
of execution processes multiple elements of the grid. This is a standard
transformation called \emph{thread coarsening}, but which we have adapted to
this specific problem. This allows some overheads to be amortized across
multiple grid elements, thus increasing performance.

\section{Background}
\subsection{Graphics Processing Units}
While originally designed for computer graphics, GPUs have become a common and
accessible approach to accelerating general-purpose computations.
Here we provide only a brief introduction to GPU architecture; a complete
discussion is beyond the scope of this paper. Two common APIs
used to program GPUs are CUDA (a proprietary standard from NVIDIA), and OpenCL
(a cross-vendor standard that is also applicable to CPUs and FPGAs). We will
use the OpenCL terminology as it is more generic, although our implementation
runs on both CUDA and OpenCL. For readers more familiar with CUDA, substitute
thread for work-item, thread-block for work-group, grid for kernel-instance,
shared memory for local memory, and streaming multiprocessor for compute
unit.

OpenCL works on a single-program multiple-data model. A single program,
called a \emph{kernel}, is executed many times in parallel. Each execution is a
\emph{work-item}. Work-items are arranged into \emph{work-groups}. The
work-items of a work-group are guaranteed to execute concurrently, and can
synchronize and communicate with each other. The set of all work-items
launched at one time is called a \emph{kernel-instance}. GPUs comprise
multiple \emph{compute units} which operate largely independently, each with
their own schedulers, L1 caches, register file and execution units --- similar
to CPU cores. Each work-group is assigned to one compute unit, but a compute
unit can run multiple work-groups concurrently.

GPUs also have multiple memory systems. The slowest, largest memory is
\emph{global memory}, which is generally off-chip DRAM. There are usually also
several levels of cache for this global memory. \emph{Local memory} is fast
on-chip memory local to a compute unit, which can be used for work-items in a
work-group to communicate with each other, and is also used as a
software-managed cache. The fastest memory is registers,
which are local to a work-item. There are other special-purpose memory
types, but they are not relevant here.

\subsection{Convolutional Gridding}
Consider the full-Sky radio interferometry measurement equation (RIME)
\citep[eq 17]{rime}:
\begin{equation}
\begin{aligned}
  K_{pq} &= e^{-2\pi i(u_{pq}l + v_{pq}m + w_{pq}(n-1))}\\
  V_{pq} &= G_p\left(\iint_{lm} \frac{1}{n}K_{pq}E_pBE_q^H \,dl\,dm\right)G_q^H.
\end{aligned}
\end{equation}
Here, $l, m, n$ are direction cosines parameterizing the sky, $(u_{pq}, v_{pq},
w_{pq})$ is the baseline vector between antennas $p$ and $q$, $B$ is the
brightness matrix at $(l, m, n)$, $E_p$ is a Jones matrix for
direction-dependent effects, $G_p$ is a Jones matrix for direction-independent
effects, and $V_{pq}$ is the predicted visibility.

With the exception of the $w_{pq}(n-1)$ term in the exponent, this is a
Fourier transform relationship between visibilities and the sky. Evaluating or
inverting the RIME directly is prohibitively expensive, so it is typically
done using fast Fourier transforms (FFTs) \citep{fft}. However,
visibilities are not sampled on a regular grid, so an extra \emph{gridding}
step must be taken to generate such a grid before using the FFT to produce an
image.

Simply snapping each visibility sample to the nearest point on the grid would
cause severe artefacts, particularly aliasing. Instead, each visibility sample
is treated as a Dirac delta, convolved with some function, and then sampled
onto the grid. Convolution in visibility space is equivalent to multiplication
in image space, so using a function with bounded support in image space
provides antialiasing \citep{vla131}. The $e^{w_{pq}(n-1)}$ and $E_p$ terms can also be
handled by convolution in visibility space --- these are known as
W-projection \citep{wprojection} and A-projection \citep{aprojection}
respectively.

The gridding convolution function (GCF) cannot always be computed
analytically, and even when it can, it is usually expensive to do so. Thus,
tables of GCFs are normally precomputed numerically. To reduce aliasing, the
GCF needs to be sampled at a higher resolution than the grid itself. A typical
value is $8\times$ oversampling \citep{romein-gridding}, but this will depend
on how far from the field of view one expects to find contaminating signals.

Efficient gridding on a GPU is challenging because the problem has irregular structure,
with the memory accesses depending on the $uvw$ coordinates. There is plenty of
parallelism, but multiple visibilities will contribute to each grid point and
so there are data hazards. A na\"\i{}ve implementation will also be
totally memory-bound: multiplying two single-precision complex numbers and
accumulating the result into memory requires 8 flops and 16 bytes of memory
traffic, while typical desktop GPUs can have compute-to-bandwidth ratios of
15--20 flops per byte.

\citet{romein-gridding} introduced the first reasonably efficient
GPU-accelerated gridding algorithm. It takes advantage of the spatial
coherence of the
data to reduce memory bandwidth. For a single baseline and frequency, the
UV-plane positions move slowly over time as the Earth rotates. Similarly,
moving to an adjacent frequency bin involves a small shift in the UV plane.
Thus, if one iterates over the visibilities for a single baseline, the GCF
footprints will almost entirely overlap. This makes it possible to maintain
sums in registers which are only occasionally flushed to global memory.

\begin{figure}
  \centering
  \begin{tikzpicture}[x=0.3cm, y=0.3cm]
  \foreach \x in {0, 8}
    \foreach \y in {0, 8}
    {
      \begin{scope}[shift={(\x, \y)}]
        \fill[shift={(4,0)},gray!50!white] (0, 0) rectangle (4, 4);
        \node[shift={(4,2)},circle,inner sep=0pt, minimum size=0.15cm,fill=black]
            at (0.5, 0.5) {};
        \node[shift={(6,1)},inner sep=0pt, minimum size=0.1cm,cross out,very thick,draw=black]
            at (0.5, 0.5) {};
      \end{scope}
    }
  \draw[help lines] (0, 0) grid[step=1] (16, 16);
  \draw[very thick] (0, 0) grid[step=8] (16, 16);
  \draw[very thick, dashed] (2, 2) rectangle (10, 10);
\end{tikzpicture}
  \caption{Overview of Romein's gridding algorithm. The dashed box shows a
    bounding box containing the GCF footprint. One work-item handles
    grid points marked with a dot; another handles those marked with a cross,
    and so on. The gray box indicates a tile: once all the grid points in a
    tile have been handled, the same work-items are recycled to update the
    next tile.
  }\label{fig:bin-orig}
\end{figure}
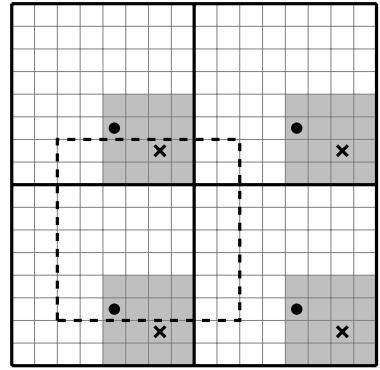

Figure~\ref{fig:bin-orig} shows how the algorithm works. The grid is divided
into \emph{bins}, which are at least as large as the GCF --- in the
original algorithm, they are the same size. A work-item is responsible for all
the positions in the grid that have the same relative placement within a bin,
e.g., all the grid positions marked with a dot are the responsibility of one
work-item. A bin-sized bounding box is placed around the GCF
footprint for one visibility, which will contain exactly one grid-point per work-item. Each
work-item maintains an in-register accumulator for that grid point. When the
bounding box moves, some work-items will switch to a different grid point:
when this happens, those work-items flush their accumulator to global memory
using an atomic addition. If the bounding box moves by one grid point, then only
$O(N)$ atomic updates are made for an $N\times N$ GCF, thus greatly
reducing the memory traffic.

Coarse-grained parallelism is achieved by assigning each baseline to a
separate work-group. Because these work-groups operate independently, they may
potentially update the same grid points at the same time; this is why grid
updates are done using atomic instructions.

A complication arises if the bins are too large to hold an entire bin in
registers at once. In this case, each bin is split into \emph{tiles}
(Figure~\ref{fig:bin-orig} shows one tile in gray), and a work-group handles
only one tile's-worth of work-items. Romein iterates serially over tiles
within the GPU code: after a work-group has iterated over all visibilities in
its baseline, it iterates over them again, but taking responsibility for the
next tile. Our implementation is parallel rather than serial, using a separate
work-group per tile. In either case, the number of atomic updates to the grid
is unaffected by tile size, but visibilities and their coordinates are loaded
from memory once for each tile in a bin.

\citet{muscat-gridding} noticed that it is not necessary to grid each
visibility individually. In some cases, particularly for short baselines, two
adjacent visibilities have the same position on the higher-resolution
grid used to sample the GCF. This means that they will be multiplied by the
same GCF samples, and thus they can be added together to form a single
visibility. This yields identical results (up to floating-point precision) but
reduces the number of visibilities to grid. He refers to this merging process
as \emph{compression}. We use compression in our implementation, and in our
results we consider only the rate for gridding these compressed visibilities,
rather than the original visibilities.

\section{Thread Coarsening}
Thread coarsening is the process of merging multiple work-items (also known as
threads) into one. This is similar to loop unrolling, but applied across
parallel work-items rather than across serial loop iterations. This improves
instruction-level parallelism \citep{volkov-benchmark}, reduces the number
of memory-access instructions \citep{gpgpu-compiler} and eliminates redundant
calculations when the same value is computed in every work-item
\citep{automatic-coarsening}.

Thread coarsening also has several negative effects. It reduces the total
amount of parallelism, which can harm performance if there is insufficient
remaining parallelism to saturate the GPU. It increases the number of
registers used per work-item, which in turn reduces the number of work-items
that can execute in parallel (\emph{occupancy}). All else being equal,
reducing occupancy will reduce latency-hiding, but with thread-coarsening it
is compensated by the increase in instruction-level parallelism
\citep{occupancy}. Finally, it can modify memory access patterns such that an
access by a sub-group is no longer to contiguous memory (so-called
\emph{coalesced} access), thus requiring more transactions with the memory
system.

Thread-coarsening can be automated by a compiler, but for gridding this will
not achieve the full benefits. The address calculations in Romein's algorithm
are different for each work-item, so thread-coarsening would not reduce the
number of instructions required. However, adjacent work-items
mostly access adjacent memory locations --- the exception being when
they are at the edge of the bounding box, causing wrap-around. We can
eliminate this case by having the bounding box move in larger steps, so that
it is always aligned to a coarser grid. This allows most of the work in
address calculation to be amortized across multiple grid points.

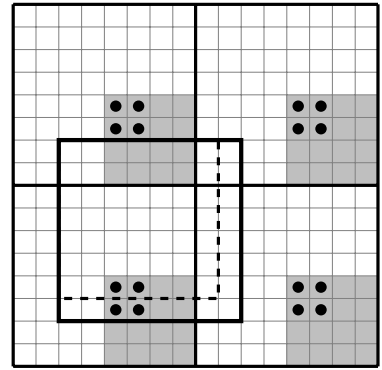
\begin{figure}
  \[\begin{tikzpicture}[x=0.3cm, y=0.3cm]
  \foreach \x in {0, 8}
    \foreach \y in {0, 8}
    {
      \begin{scope}[shift={(\x, \y)}]
        \fill[shift={(4,0)},gray!50!white] (0, 0) rectangle (4, 4);
        \foreach \i in {4,5}
          \foreach \j in {2,3}
          {
            \node[shift={(\i,\j)},circle,inner sep=0pt, minimum size=0.15cm,fill=black]
               at (0.5, 0.5) {};
          }
      \end{scope}
    }
  \draw[help lines] (0, 0) grid[step=1] (16, 16);
  \draw[very thick] (0, 0) grid[step=8] (16, 16);
  \draw[ultra thick] (2, 2) rectangle (10, 10);
  \draw[very thick,dashed] (2, 3) rectangle (9, 10);
\end{tikzpicture}\]
  \caption{Mapping of work-items and work-groups to grid points. One work-group
   contributes to all the shaded cells. One work-item contributes to all the
   cells marked with a dot. Here blocks are $2\times 2$, tiles are
   $4\times 4$ and bins are $8\times 8$. The dashed box shows a GCF
   footprint, and the solid box shows the bounding rectangle.
  }\label{fig:bin}
\end{figure}

Figure~\ref{fig:bin} shows how this is implemented. Each
work-item now handles a \emph{block} of grid points ($2\times 2$ in the
figure), and the bounding box is aligned to the edges of blocks. Of course,
the footprint of the GCF is unaffected, so the bounding box must now be
slightly larger than the GCF to ensure that a suitably aligned bounding box
can always be found to contain the GCF. Grid-point updates still occur for the grid
points in the padding between the GCF footprint and the bounding box, so
the storage for the GCF must be padded with zeros. These updates are also
wasted computations, but provided the blocks are significantly smaller than
the GCF, this will add only a small amount of overhead.

With this change, expensive address calculations that were previously done
per grid point are now only done once per block. In particular, it is only
necessary to check whether the whole block has moved out of the bounding box,
rather than each grid point. This reduces the total number of instructions
needed for addressing and thus frees up more cycles for actual gridding
calculations.

\section{Implementation Details}
\citet{muscat-gridding} implements compression on the fly during gridding. For
deconvolution with major cycles \citep{cotton-schwab}, the compression only
needs to be done once for all cycles, so we have implemented it as a
preprocess. For each compressed visibility, the gridding kernel receives the
integer grid coordinates, the sub-grid coordinates for indexing the GCF, the
$w$ plane index, and the pre-weighted visibility values. Rather than all
work-items directly loading these values from global memory, they are staged
via local memory in batches. When loading a batch, each work-item loads one
visibility.

We found that the NVIDIA CUDA compiler was causing the kernels to use far more
registers than we expected. Examining the assembly code, we found that
it was using extra scratch registers for flushing accumulators. The CUDA C
code would first atomically add the accumulator to global memory, then set
it to zero. The assembly code instead copied the accumulator to a temporary,
zeroed the accumulator, then atomically added the temporary to global memory.
This was presumably done to improve latency-hiding (allowing the accumulator
to be re-used before the atomic operation completed). However, this increases
the number of 32-bit registers by two times the number of polarizations times
the coarsening factor. In this case, the extra register pressure reduced
occupancy so much that performance dropped overall. As a workaround, we added
the statement \lstinline'asm("")' as a compiler-level memory barrier.

\citet{romein-gridding} found that the majority of memory traffic was due to
cache misses in reading the convolution GCF values. To avoid this problem,
we have used a separable approximation to the GCF \citep{separable-w}.
We have also (like Romein) assumed that the GCF is
polarization-independent. Both these assumptions mean that our results cannot
be directly applied to A-projection; we nevertheless expect our technique to
provide similar accelerations for A-projection, provided the memory system can
keep up.

Coarsening also works nicely with a separable GCF. For an $m\times n$ block,
we load $m + n$ values from memory in each work-item, and compute the $m\times
n$ products in registers. Larger blocks thus reduce the number of memory
transactions required.

It is difficult to determine the best coarsening factors and work-group size
theoretically, because there are trade-offs. For example, a larger work-group
implies fewer tiles per bin, and thus less memory traffic to load
visibilities; but only an integer number of work-groups can be active on a
compute unit at a time, so larger work-groups may cause resources to be
under-utilised. We have dealt with this by using autotuning: a small
artificial data-set is synthesized and benchmarked with a range of coarsening
factors and work-group sizes, and the best combination is remembered. We
perform auto-tuning separately for each number of polarizations.

Another tuning factor is the number of visibilities to process in each
work-group. \citet{romein-gridding} uses one work-group per baseline; but with
compression, that will lead to unbalanced workloads. We order visibilities by
baseline, but assign a fixed number of visibilities to each work-group. Larger
numbers make more use of spatial coherence between adjacent visibilities, but
reduce parallelism. Our current implementation uses \num{1024} per work-group.

\section{Results}

We found that the performance of our gridding implementation is highly
dependent on the coordinates and ordering of the data. To obtain a dataset
that is representative of future radio telescopes, we simulates a
single-channel, two-hour observation on MeerKAT \citep{meerkat-params} with a
\SI{2}{\second} integration time, and compressed the visibilities. These
compressed visibilities are available online
\footnote{\url{https://github.com/ska-sa/thread-coarsening-grid-data}} for anyone who wishes
to do a direct comparison. All gridding calculations are done in
single-precision floating point.

\begin{figure}
  \centering\includegraphics{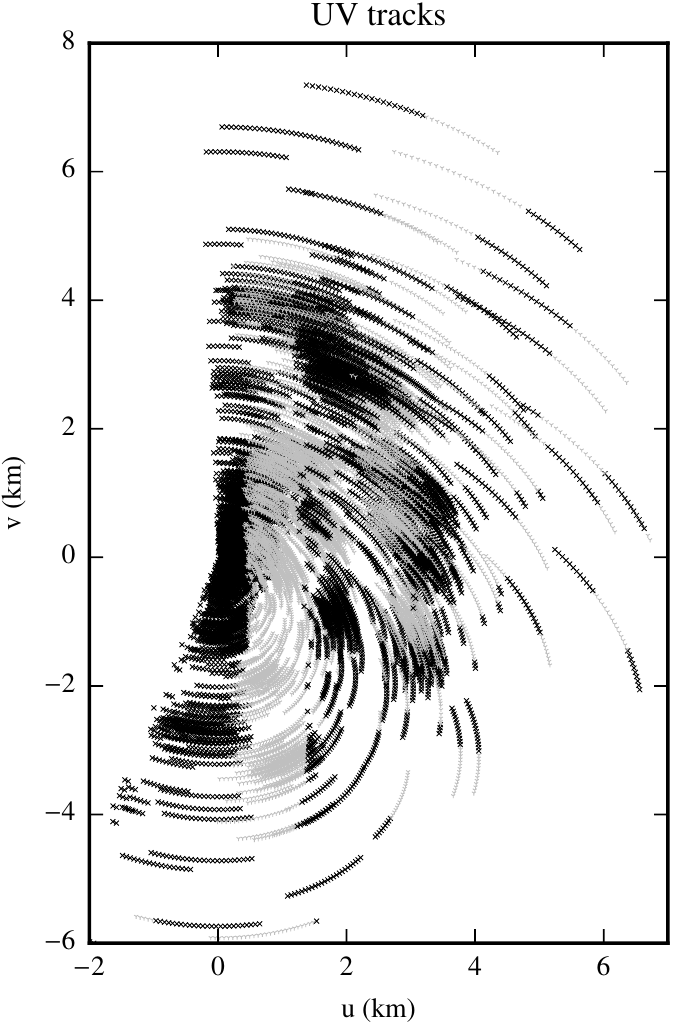}
  \caption{The $uv$ coordinates in our simulation. To improve legibility, only
    every 100th visibility is shown. The alternating bands of black and gray
    correspond to the slices used for W-stacking (they overlap because the
    array is not perfectly coplanar).
  }
  \label{fig:track}
\end{figure}

Pure W-projection requires large GCFs to correct the W
effects, making it very slow. For MeerKAT, we expect it to be used in
conjunction with techniques such as W-stacking \citep{wsclean}. We have thus
split the data into a number of slices by $w$ value, as shown in
Figure~\ref{fig:track}. We invoke a separate kernel-instance for each slice
in the stack. There are 7 slices
with a total of \num{2337867} compressed visibilities (from \num{7257600}
uncompressed), and \num{1229} W planes per slice. Note that in real use, the
number of slices would be adapted to the size of the GCF (or vice versa),
but we have kept the number of slices fixed while varying the GCF size so
that we can study the effect of GCF size in isolation.

It should be emphasized that these choices are largely conservative. The use
of such a short observation with a single channel, which is further split into
W slices, substantially reduces parallelism. The results reported here are
thus likely to be achievable in other configurations.

We explored a variety of configurations: coarsening factor of 1, 2, 4 or 8 on
each axis (up to a combined factor of 16), work-group size of 4, 8 or 16 on
each axis, bin sizes of 32, 64 or 128; and either 1 or 4 polarizations. The
GCF size was computed as
$\text{bin size} + 1 - \max(\text{coarsen}_u, \text{coarsen}_v)$, which is the
largest square size possible.

We report results on two GPUs: an NVIDIA GTX
980 (Maxwell architecture) using CUDA, and an AMD Radeon R9 290X (GCN
architecture) using OpenCL. These have theoretical single-precision
performance of \SI{5.288}{\tera flop\per\second} and \SI{5.632}{\tera
flop\per\second}, and memory bandwidth of
\SI{224}{\giga\byte\per\second} and \SI{352}{\giga\byte\per\second}
respectively.
The implementation was developed and tuned on the
Maxwell architecture, so we give the most attention to results on the GTX 980,
and results are for this GPU except where otherwise noted.
The R9 290X is included to show that the optimizations are not specific to one
GPU architecture or API.

\begin{figure}
  \centering\includegraphics{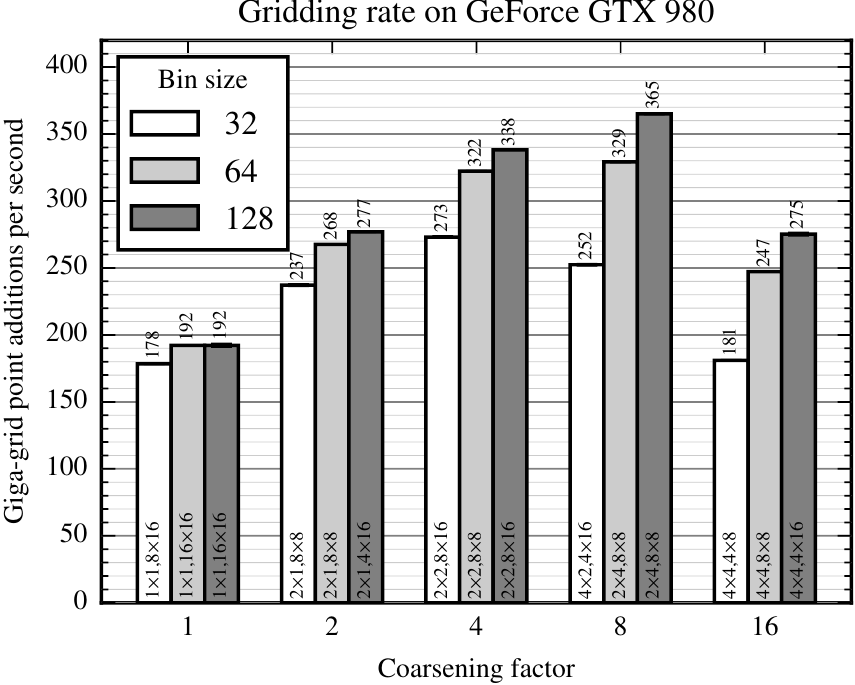}
  \caption{Gridding rates for four polarizations (GTX 980). The numbers inside
    the bars are the coarsening factors (in U and V), and the number of
    work-items per work-group (in U and V) in the best case.}
  \label{fig:grid_p4_gtx980}
\end{figure}

Figure~\ref{fig:grid_p4_gtx980} shows gridding rates for four polarizations.
For each total coarsening factor, we show only the result for
the best combination of U and V coarsening factors and work-group size.
We consider a grid-point addition to be an addition for a
single polarization, which requires 8 flops. The efficiency without
coarsening is 27--29\%, which is similar to (but slightly higher than) the
24\% reported by \citet{romein-gridding}, and much higher than the results of
\citet{muscat-gridding}.

In the best case (bin size 128, coarsening factor 8) we achieve \SI{365}{\giga
GPA\per\second}, at 55\% efficiency. This is a 90\% improvement over
no coarsening at the same bin size. For a bin size of 32, the
improvement due to coarsening is only 53\%. Improvements will be less at
smaller bin sizes because too much coarsening will quickly increase the
proportion of flops wasted by padding and reduce the available parallelism.
This can be seen in the performance drop from $4\times$ to $8\times$
coarsening.

\begin{figure}
  \centering\includegraphics{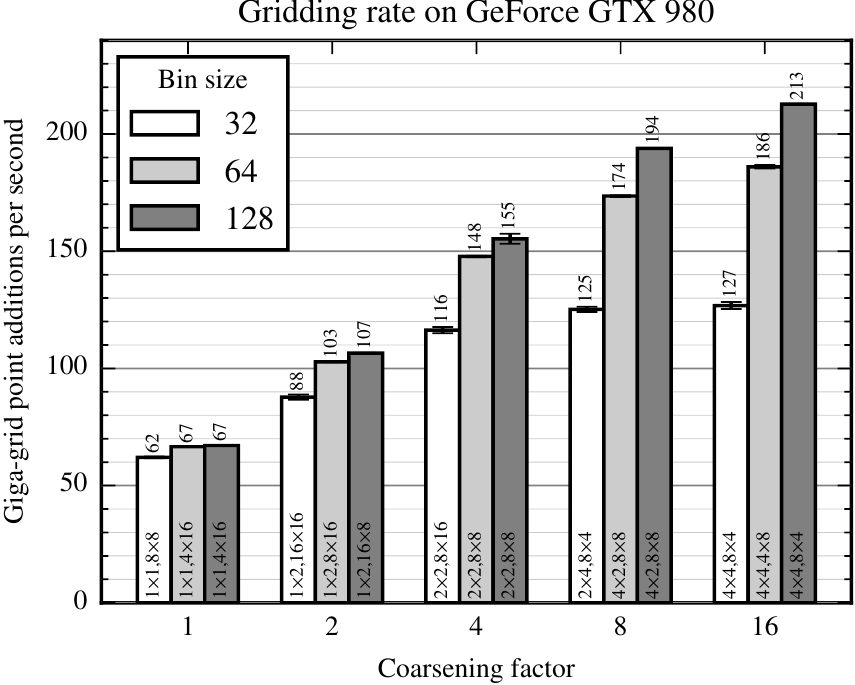}
  \caption{Gridding rates for a single polarization (GTX 980).}
  \label{fig:grid_p1_gtx980}
\end{figure}

Figure~\ref{fig:grid_p1_gtx980} shows gridding rates for a single polarization. Here,
the ratio of address calculations to gridding calculations is four times
larger, and thread coarsening makes a larger impact. For
128-pixel bins, the performance improves by a factor of $3.2$.
We also tested with larger coarsening factors, but they performed worse than
those shown.

\begin{figure}
  \centering\includegraphics{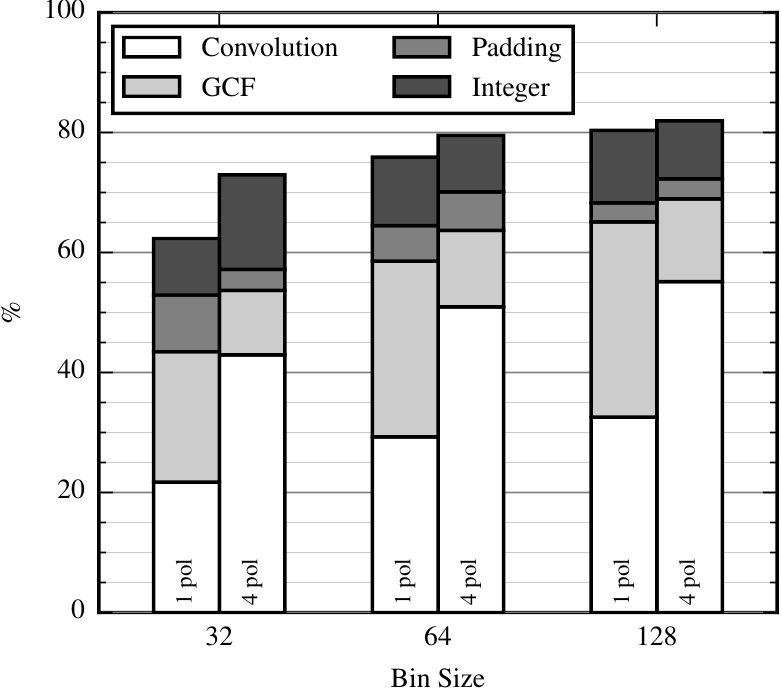}
  \caption{Utilization of the single-precision units. \emph{Convolution}:
    instructions that multiply GCF samples with visibilities and accumulate the
    result. \emph{GCF}: instructions to reconstruct GCF samples from the
    separable parts. \emph{Padding}: flops that are wasted due to padding of
    the GCF.
  }
  \label{fig:time}
\end{figure}

Figure~\ref{fig:time} shows the utilization of the single-precision units in
the GPU, which handle both floating-point and integer instructions. This shows
that although utilization is not much higher for four polarizations than for
one, the proportion of instructions used for convolution is much higher.
The lower utilization for smaller bin sizes is largely due to a
lack of parallelism in the more sparsely populated W slices --- in the worst
case, there are not enough work-groups to give every compute unit work to do.
If we consider only the first slice, then utilization levels are all between
75\% and 85\%, and the NVIDIA profiler considers this to be compute-bound.

The profiler also reports bandwidths and utilization for the various memory
systems. Table~\ref{tbl:bandwidths} shows the bandwidths for the best case.
The decline in bandwidth from the unified cache (L1) to L2 to device (global)
memory shows that the caches are effective.
\begin{table}
  \caption{Memory bandwidths for the best case (four polarizations, 128-pixel
    bins, $8\times$ coarsening), on GTX 980.
  }\label{tbl:bandwidths}
  \begin{filecontents*}{bandwidth-best.csv}
Memory Type,Read,Write
Shared,2.92357443511e+11,9137045389.47
Unified cache,1.25542769696e+12,0.0
L2 cache,1.17165765018e+11,40071748681.1
Device,14522297959.4,6354190029.41
\end{filecontents*}
\pgfplotstabletypeset[
    col sep=comma,
    columns/Memory Type/.style={string type, column type={l},},
    columns/Read/.style={
      divide by=1e9, fixed, fixed zerofill, precision=1, dec sep align,
      column name={Read (\si{\giga\byte\per\second})}
    },
    columns/Write/.style={
      divide by=1e9, fixed, fixed zerofill, precision=1, dec sep align,
      column name={Write (\si{\giga\byte\per\second})}
    }
  ]{bandwidth-best.csv}
\end{table}

We now report results for the Radeon R9 290X.
Since OpenCL doesn't support atomic additions, we had to emulate them using
slower compare-and-swap operations. Since these operations are not affected by
our optimizations, we expect thread coarsening to have less impact.

\begin{figure}
  \centering\includegraphics{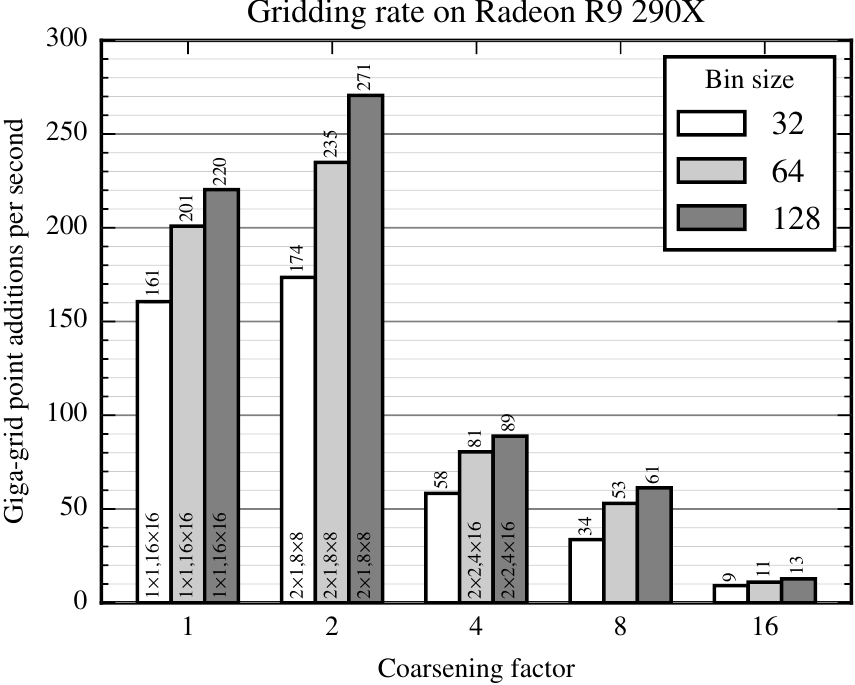}
  \caption{Gridding rates for four polarizations (R9 290X).}
  \label{fig:grid_p4_r9-290x}
\end{figure}
\begin{figure}
  \centering\includegraphics{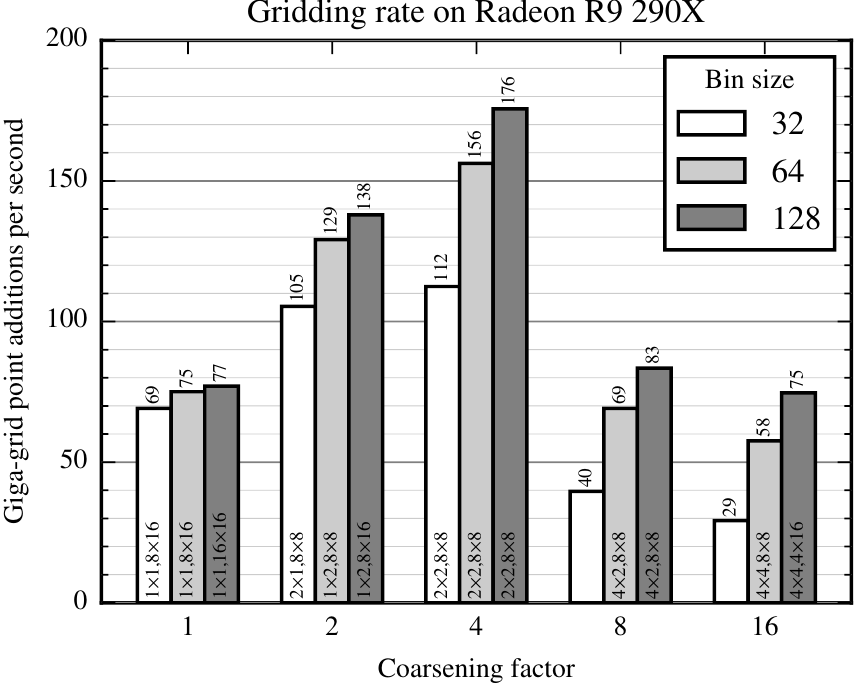}
  \caption{Gridding rates for a single polarization (R9 290X).}
  \label{fig:grid_p1_r9-290x}
\end{figure}

Figures~\ref{fig:grid_p4_r9-290x} and \ref{fig:grid_p1_r9-290x} show the
speedups obtained with thread coarsening on this GPU. One noticeable
difference from the GTX 980 figures is that the optimal coarsening factor is
lower. This suggests
that register pressure is an issue, but we have not investigated whether this
is due to compiler quirks such as the one we worked around for the NVIDIA
compiler. Between the emulated atomic additions and this inability to utilize
higher coarsening factors, it is not surprising that coarsening is less
effective than on the GTX 980. Nevertheless, it is beneficial, with up to 23\%
improvement for four polarizations and 128\% improvement for one polarization.

\section{Conclusions and Future Work}

Thread coarsening clearly provides a significant performance improvement, and
we recommend that gridders based on Romein's algorithm should use
thread coarsening to improve performance, unless they use only very small
GCFs (for example, because W effects are corrected by other means than W
projection).

We have shown significant gains in all the tested configurations, but the
largest gains are for large GCFs. The results for small GCFs are hurt by a
lack of parallelism in the sparse W slices, which could be improved by
adapting tuning parameters (both the coarsening factor and the number of
visibilities per work-group) to the number of visibilities.

For large GCFs, performance is limited by the single-precision units, so
further significant optimizations can only come from reducing the number of
flops. For each visibility $V$, polarization $p$ and position $i,j$ in the
GCF footprint, we need to compute the product
$G_{ij}V_p = (G^{\text{u}}_i G^{\text{v}}_j)V_p$. We currently compute this
as written, but \citet{separable-w} notes that fewer flops are required to
compute this as $G^{\text{u}}_i (G^{\text{v}}_j V_p)$ because the factor in
parentheses is independent of $i$. For a bin size $B$ and number of
polarizations $P$, this reduces the number of required multiplications from
$B^2 + B^2P$ to $BP + B^2P$. This is not an entirely free optimization,
because the common factor needs to be broadcast to all work-items that need
it, but we expect that performance should still improve.

Autotuning is clearly important to find the optimum coarsening factor. Our
results show that tuning needs to consider both the number of polarizations
and the bin size. On the other hand, one need only consider coarsening shapes
that are as square as possible i.e., in $1:1$ or $2:1$ ratio. This is expected,
because squarer shapes reduce both the amount of zero padding and the number
of memory accesses to load the separable GCF components.

Our implementation uses power-of-two sizes for bins, because this simplifies
some of the integer arithmetic in coordinate calculations. However, this is
not a fundamental limitation, and it would be possible to use non-power-of-two
sizes at all levels of the hierarchy. This may yield further small performance
improvements by providing a finer-grained set of options.

A natural next step is to apply thread coarsening to degridding in a similar
fashion. Our preliminary results show that performance improvements are even
better than for gridding, but we have not yet fully optimized our degridder
and this may affect the results. Nevertheless, we expect thread coarsening to
be important in degridding.

We have not considered A-projection in our performance measurements.
A-projection requires significantly more instructions and bandwidth for
loading the GCFs, because they are dependent on time, frequency, polarization
and possibly baseline, and are not separable. We thus expect thread-coarsening
to have a smaller effect, because bandwidth and latency considerations will
play a larger part.

\section*{References}

\bibliography{thread-coarsening-grid}

\end{document}